\documentclass[journal=jctcce,manuscript=article]{achemso}

\usepackage[version=3]{mhchem} 
\usepackage{amsmath}
\usepackage{amsfonts}

\title{IceCoder: Identification of Ice phases in molecular simulation using variational autoencoder}

\author{Dibyendu Maity}
\author{Suman Chakrabarty}
\email{sumanc@bose.res.in}
\affiliation{Department of Chemical and Biological Sciences,
S.\ N.\ Bose National Centre for Basic Sciences, Kolkata 700106, India}

\begin{document}

\begin{abstract}
The identification and classification of different phases of ice within molecular simulations is a challenging task due to the complex and varied phase space of ice, which includes numerous crystalline and amorphous forms. Traditional order parameters often struggle to differentiate between these phases, especially under conditions of thermal fluctuations. In this work, we present a novel machine learning-based framework, \textit{IceCoder}, which combines a variational autoencoder (VAE) with the Smooth Overlap of Atomic Positions (SOAP) descriptor to classify a large number of ice phases effectively. Our approach compresses high-dimensional SOAP vectors into a two-dimensional latent space using VAE, facilitating the visualization and distinction of various ice phases. We trained the model on a comprehensive dataset generated through molecular dynamics (MD) simulations and demonstrated its ability to accurately detect various phases of crystalline ice as well as liquid water at the molecular level. IceCoder provides a robust and generalizable tool for tracking ice phase transitions in simulations, overcoming limitations of traditional methods. This approach may be generalized to detect polymorphs in other molecular crystals as well, leading to new insights into the microscopic mechanisms underlying nucleation, growth, and phase transitions, while maintaining computational efficiency.
\end{abstract}

\section{Introduction}
Ice exhibits a remarkable diversity of polymorphs with more than twenty crystalline phases and three amorphous forms\cite{Zheligovskaya2006,BartelsRausch2012, Fuentes-Landete}. Theoretical models predict the existence of even more phases, most of which remain undiscovered yet\cite{Zhu2020}. Hexagonal ice (Ice-Ih) is the most common form that occurs naturally\cite{Baker2019,Murray_2006,Kamb1964,Hasted_1969,Bauer_2008,Kuhs_1984,Buffett2004, Buchanan2005, Moon2003,B500373C,  Johari2007, Stroeve2014, serreze2003record}. There are various ice phases that appear under certain thermodynamic conditions, ranging from conditions out in space (Ice II) to sea floor (clathrates). The rich and complex phase diagram made up of a wide variety of ice phases poses difficulties for molecular simulations, particularly in the context of detecting a specific ice polymorph in a simulation trajectory\cite{Hansen2021, Chaplin2019, Bore2022, Hirata2017}. In order to understand the molecular mechanisms underlying various phenomena, such as phase transitions, nucleation, and growth of specific polymorphs, it is essential to develop effective order parameters that can locally distinguish between transient phases emerging during simulations.

\begin{figure*}[!hbt]
\includegraphics*[width=.95\linewidth]{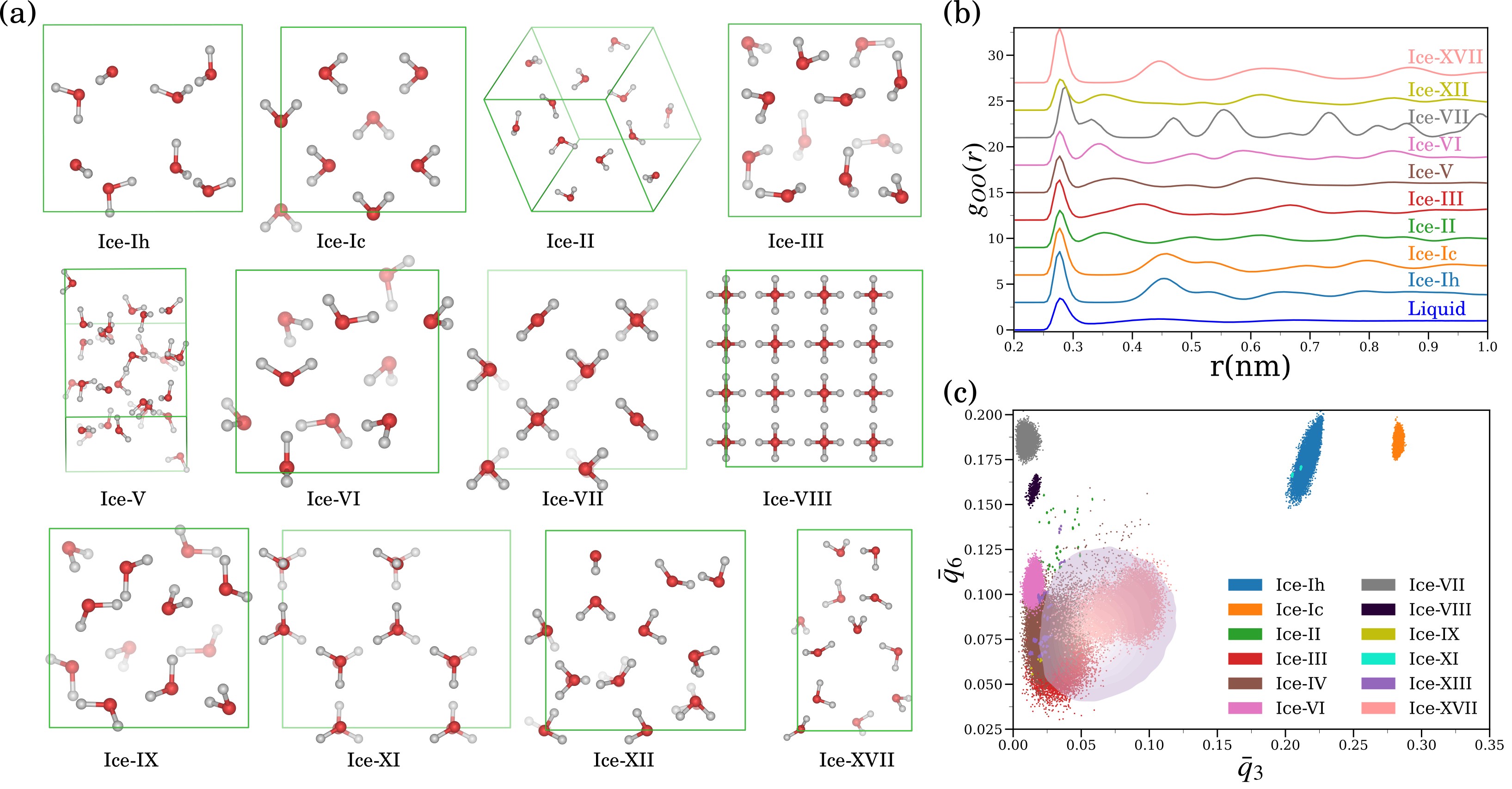}
\caption{(a) Few of the representative ice phase unit cells shown in ball and stick representation. They all are different geometrically.(b) Oxygen-oxygen pair correlation function(RDF) of the ice phases has been plotted. The difference in RDF speaks of the different atomic neighborhood of the phases.(c)A 2-dimensional scatter plot of bond orientational order parameters, $\bar{q}_3$-vs $\bar{q}$ is shown. Each point represents orientation of each oxygen atom in the specific ice phase. While some phases are clearly distinguished, significant overlap is observed among others within this parameter space.}
\label{fig:intro}
\end{figure*}

Order parameters are mathematical functions based on atomic coordinates that can differentiate between specific states or phases of matter. Their design often relies on intuition and the underlying physics of the problem, which is a challenging task. Several physics-based order parameters have been developed to distinguish between phases considering atomic geometry, orientation, or neighborhood arrangement\cite{CHAU_1998, Giovambattista_2005, Yan_2007, Zimmermann2017, Errington2001, Piaggi2017, Lechner2008, Reinhardt2012, Mickel2013, Crippa2023}. 

For example, the tetrahedral order parameter (Q), has been used to quantify local structuring (tetrahedrality) in bulk water, interfacial water and water under confinement\cite{Errington2001, CHAU_1998, Galamba_2013, Giovambattista_2005, Yan_2007, Hande2015, Hande2016, Hande2022}. Four-body structural order parameter ($F_4$) has been successful in distinguishing between regular ice, liquid, and clathrate structures\cite{Rodger1996,Choudhary2018, Choudhary2020}. The Local Structure Index (LSI) has proven useful in studying supercooled water, protein hydration shells, and water near hydrophobic surfaces\cite{Shiratani1998, Wikfeldt2011}. Recently, entropy fingerprints, particularly their local variant (local entropy), utilizing the pair correlation function of oxygen atoms in water molecules, have been used to differentiate hexagonal and cubic ice from water efficiently\cite{Piaggi2017}. 

Steinhardt order parameters, which leverage the orientation of neighboring molecules and achieve roto-translational invariance through spherical harmonics, have successfully identified regular lattice structures in simulations\cite{Plimpton1995}. In the context of ice, the $q_3$ and $q_4$ variant of Steinhardt's parameter has been applied to explore the ice phase diagram\cite{Doi2021, Kim2020}. The CHILL\cite{Moore2010} and CHILL+\cite{Nguyen2014} algorithms, based on Steinhardt’s parameters, have been developed to quantify the number of ice phases in a simulation, capable of detecting hexagonal, cubic, and clathrate phases in liquid water.

However, the challenge lies in the subtle distinctions between these ice phases, with differences so minor that conventional geometric methods may fail to differentiate them. Often, specific order parameters are tailored for individual case studies, yet they may not be universally applicable across all ice phases. In addition, these methods frequently require substantial computational resources, limiting their broader application. In response to these challenges, machine learning (ML) methods have gained popularity as a promising approach to formulating better order parameters\cite{ML_Method1, Cheng2019, Hofstetter2022, Kim2020}. Using extensive data sets, ML algorithms can uncover complex relationships within intricate data, revealing patterns that might otherwise remain hidden.

In this context, Geiger and Dellago introduced a classification method based on supervised feed-forward neural networks\cite{Geiger2013}. Using symmetry functions to characterize the atomic neighborhood, they successfully classified multiple ice phases, including ice-Ih, ice-Ic, ice-II, ice-III, and ice-V. Building on this, Fulford \textit{et al.} have developed the DeepIce network, which utilizes compact subnetworks that take the Cartesian and spherical coordinates of water molecules as input\cite{Fulford2019}. This supervised approach can accurately distinguish between ice-Ih, ice-Ic, and liquid phases without relying on prior knowledge of molecular structures, offering greater precision in comparison to methods based on Steinhardt parameters.

Subsequently, DeFever et al. proposed a PointNet architecture that directly processes point clouds of the system without any preprocessing\cite{DeFever2019}. This approach extended the classification capabilities to include additional phases, such as ice-VI and hydrates, alongside those identified by Geiger and Dellago. Another contribution is GCIceNet by QHwan Kim et al., a deep learning-based model for ice phase classification that incorporates convolutional neural networks (CNNs) into water studies for the first time\cite{Kim2020}. GCIceNet considers plastic ice and high-pressure ice phases in addition to the aforementioned phases and offers accuracy of the order of DeepIce and PointNet, although it still encounters overlaps between states, particularly in the liquid regime. Similarly, there has been a concurrent effort to classify crystal structures derived from liquid simulations. In this regard, recent work by Zou and Tiwary demonstrated that a graph neural network (GNN)-based autoencoder model can effectively differentiate between various crystal structures in simulations at a local level and take advantage of latent representations towards enhanced sampling of rare events\cite{Zou2024}. Kuroshima \textit{et al.} presented a comparative analysis of a GNN-based classifier and a classifier utilizing molecular symmetry functions, using t-SNE embedding for visualization\cite{Kuroshima2024}.

Although there exists a plethora of methodologies to classify ice phases, the challenge of having a robust yet general low-dimensional order parameter to classify and distinguish all known ice phases remains elusive. In this study, we first evaluated the performance of traditional order parameters for classifying a large number of ice phases to demonstrate that none of them are good enough to distinguish between all of the ice phases investigated here. We have developed a deep learning approach based on the variational autoencoder\cite{VAE} (VAE) combined with a high-level machine learning descriptor known as Smooth Overlap of Atomic Positions(SOAP)\cite{Himanen2020,De2016, Maksimov2020}. Our approach aims to generalize the classification of ice phases using a convenient two-dimensional order parameter space for easy visualization and interpretation of the phase transformations.

We generated an extensive dataset for each ice phase along with the liquid state using molecular dynamics (MD) simulations. We represent the atomic environments using SOAP vectors as described later. These high-dimensional vectors are then compressed using a variational autoencoder-based artificial neural network (ANN) architecture. We demonstrate that this model is capable of detecting several ice phases at a single-molecule (local descriptor) level with high efficiency. In Section II of this article, we discuss the methodologies employed to address the challenges of ice phase detection and the formulation of the order parameters used. Section III presents our results, highlighting the capabilities of our method and the limitations of traditional order parameters in classifying ice phases. Finally, the concluding section summarizes our observations. 

\section{Methods}
\subsection{A brief overview of the common order parameters}
Order parameters (OPs) are typically designed based on an intuitive understanding of the specific problem at hand. These parameters often focus on the spatial arrangement of neighboring atoms within a local environment, encoding geometric information through appropriate mathematical functions of the atomic coordinates. In the following, we discuss the formulation of a few commonly used OPs to quantify water structure.

The \textbf{tetrahedral order parameter(Q)} quantifies the degree of local ordering in a system, particularly the extent to which water molecules arrange themselves in a tetrahedral configuration. This OP has been widely used in studies of water structure, especially near hydrophobic surfaces, where water molecules form ordered structures distinct from those of bulk water \cite{Errington2001, Jedlovszky2008}. The tetrahedral order parameter is mathematically defined as:
\[
Q(i) = 1 - \frac{3}{8} \sum_{j=1}^3 \sum_{k=j+1}^4 \left(\cos\phi_{jk} + \frac{1}{3}\right)^2
\]

Here, the four closest neighbors of the water molecule are considered, with $\phi_{jk}$ representing the angle formed by the oxygen atom under consideration and its two nearest neighbors, \( j \) and \( k \).

\textbf{Steinhardt's order parameters}, a class of local bond orientational order parameters, are another widely used OPs. These are based on spherical harmonics and are often utilized to identify crystal structures in molecular simulations \cite{Plimpton1995}. They classify the phase of an atom or molecule by assessing the coherence of its orientational order with that of its neighbors, making them particularly useful for analyzing the structure of ice poly morphs, the crystalline solids characterized by a high degree of local ordering\cite{Reinhardt2012}. 

To calculate Steinhardt's order parameters, the nearest neighbors of each particle are first identified. The angles between the vectors connecting the particle to its neighbors are then calculated and used to derive rotationally invariant spherical harmonics. The Steinhardt's order parameter of degree \( l \) for the oxygen of molecule \( i \), denoted \( q_l(i) \), is defined as:
\[
q_l(i) = \sqrt{\frac{4\pi}{2l+1} \sum_{m=-l}^l |q_{lm}(i)|^2}
\]

where the \( (2l + 1) \) complex components, \( q_{lm}(i) \), are given by:
\[
q_{lm}(i) = \frac{1}{N_b(i)} \sum_{j=1}^{N_b(i)} Y_{lm}(\textbf{r}_{ij})
\]

In this expression, \( N_b(i) \) is the number of nearest neighbors within a specified cutoff distance, and \( Y_{lm}(\textbf{r}_{ij}) \) represents the spherical harmonics for the unit vector \( \textbf{r}_{ij} \) connecting the molecule \( i \) to its neighboring molecule \( j \). A more refined distinction between atomic environments can be achieved using the averaged form of \( q_l \):
\[
\bar{q}_l(i) = \sqrt{\frac{4\pi}{2l+1} \sum_{m=-l}^l |\bar{q}_{lm}(i)|^2}
\]

where the \( q_{lm} \) vector of molecule \( i \) is averaged with those of its nearest neighbors, including itself:
\[
\bar{q}_{lm}(i) = \frac{1}{N_b(i) + 1} \sum_{j=0}^{N_b(i)} q_{lm}(j)
\]

It is important to note that Steinhardt's parameters are highly sensitive to the choice of nearest-neighbor scheme and the cutoff value for the nearest neighbor search. Several methods exist for selecting the nearest neighbors of a reference atom, including fixed radius cutoff, fixed numbers of nearest neighbors, Voronoi diagrams, and their dual, the Delaunay graph \cite{Mickel2013} etc.

The concept of an \textbf{entropy fingerprint} \cite{Piaggi2017} utilizes the dominant two-body term, referred to as ``pair entropy" or ``two-body excess entropy"(\( S_2 \)) \cite{Saija2003, Truskett2000}, within the context of configurational entropy expansion in terms of multibody correlation functions\cite{Baranyai1989, Nettleton1958}. Here, only the radial pair distribution function \( g(r) \) for oxygen-oxygen interactions is considered. The pair entropy is expressed as:
\[
S_2 = -2 \pi \rho k_B \int_0^{\infty} \left[g(r) \ln(g(r)) - g(r) + 1\right] r^2 \, dr
\]

In this equation, \( \rho \) represents the density of the system, \( g(r) \) is the radial pair distribution function, and \( k_B \) is the Boltzmann constant. Although entropy is inherently a global property, it can be projected locally onto each particle \( i \) to define a local order parameter:
\[
e(i) = -2 \pi \rho k_B \int_0^{r_m} \left[g_m^i(r) \ln(g_m^i(r)) - g_m^i(r) + 1\right] r^2 \, dr
\]

To obtain a continuous and differentiable order parameter, the radial distribution function \( g^i_m(r) \) centered on particle \( i \) can be smoothed using Gaussian functions with  finite width (e.g. 0.15 Å). In this context, \( r_m \) denotes an upper integration limit, typically set to 5.0 Å, although it should approach infinity in a strict entropy definition.

For enhanced resolution between different environments, a locally averaged entropy (\( le \)) can be defined as:
\[
le(i) = \frac{1}{N_b(i) + 1} \sum_{j=0}^{N_b(i)} e(j)
\]
In this expression, the sum runs over the \( N_b(i) \) neighbors of particle \( i \), including particle \( i \) itself. The summation over the first neighbors is carried out using the same switching function employed for Steinhardt's parameters.

\subsection{Architecture and computational details of IceCoder}
IceCoder is primarily based on a variational autoencoder (VAE) \cite{VAE}, enhanced by incorporating the SOAP representation as a feature-engineered input. Like a standard autoencoder, a VAE comprises both an encoder and a decoder. The primary objective during training is to minimize the reconstruction error between the encoded-decoded data and the original input data. However, to introduce regularization into the latent space, the VAE modifies the encoding-decoding process. Instead of encoding an input as a single point, it is represented as a distribution over the latent space.

The loss function minimized during VAE training consists of two components: a reconstruction term that optimizes the encoding-decoding process on the final layer, and a regularization term that encourages a well-behaved latent space. The regularization term is defined as the Kullback-Leibler (KL) divergence between the distribution produced by the encoder and a standard Gaussian distribution \cite{Kullback1951, Belov2011}. The VAE loss function is mathematically expressed as:
\[
\begin{split}
    \text{loss}_{\text{VAE}}(\phi, \theta) = & -\sum_{i=1}^n \mathbb{E}_{z_i \sim q_\phi(z_i \mid x_i)} \left[\log p_\theta(x_i \mid z_i) \right] \\
    & - KL\left(q_\phi(z_i \mid x_i) \ || \ p(z_i)\right)
\end{split}
\]
The first term represents the reconstruction loss, which drives the model to accurately reconstruct the original input \(x_i\) from its compressed representation \(z_i\). This loss is typically calculated as the mean squared error between the original and reconstructed data:
\[
\frac{1}{n} \sum_{i=1}^n \|x_i - f_\theta(h_\phi(x_i))\|^2
\]
where \(h_\phi\) and \(f_\theta\) are the encoding and decoding neural networks, respectively.

The second term (KL-divergence) acts as a regularizer for the latent space. This term promotes the latent variables generated by \(q_\phi(z_i \mid x_i)\) to follow the prior distribution \(p(z_i)\), which is typically a standard normal distribution.

In terms of representing atomic environments, several methods have been developed, but our approach focuses on using the SOAP representation. The atomic neighbor density for a given atom is defined as:
\[
\rho_{\chi}^\alpha(\vec{r}) = \sum_i w_{z_i} \delta(\vec{r} - \vec{r_i})
\]
where the index \(i\) sums over neighboring atoms within a specified cutoff distance \(r_c\). The term \(w_{z_i}\) is a weight factor based on the atomic species of the neighboring atom \(i\), and \(\vec{r_i}\) is the vector from the central atom to the neighbor \(i\).
The delta functions \(\delta\) are then smoothed using Gaussian functions centered at the positions of the neighboring atoms. The density function is thus modified to:
\[
\rho_{\chi}^\alpha(\vec{r}) = \sum_{|\vec{r_i}| < r_c} \exp\left(-\frac{|\vec{r} - \vec{r_i}|^2}{2\sigma^2}\right)
\]
The density function($\rho_{\chi}^\alpha(\vec{r})$) is further expanded using orthonormal radial functions \(g(|\vec{r}|)\) and spherical harmonics \(Y_{lm}(\hat{r})\):
\[
\rho_{\chi}^\alpha(\vec{r}) = \sum_{nlm} c^\alpha_{nlm} g_n(|\vec{r}|) Y_{lm}(\hat{r})
\]
Finally, the power spectrum is computed as:
\[
p^\alpha_{nn'l}(\chi) = \sqrt{\frac{8}{2l+1}} \sum_m \left(c^\alpha_{nlm}\right)^* c_{n'lm}
\]
This power spectrum provides a translation, permutation, and rotation-invariant characterization of the local environment. The resulting vector, approximated by truncation beyond certain cut-off values \(l_{\text{max}}\) and \(n_{\text{max}}\), serves as the fingerprint of the local environment.

\begin{figure}
	\centering
	\includegraphics*[width=.95\linewidth]{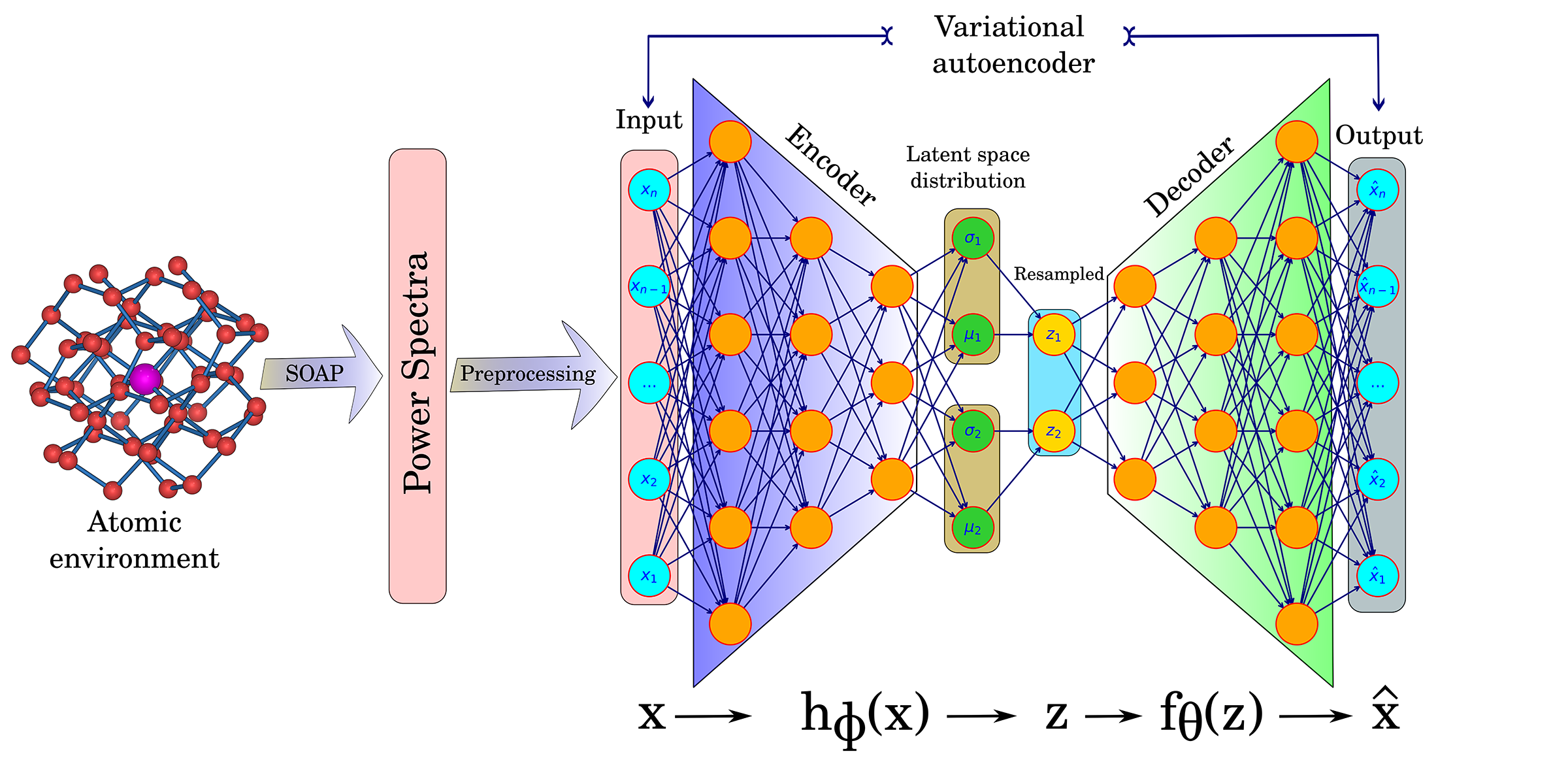}
	\caption{The model's architecture is outlined alongside a schematic of the workflow. At first, the atomic neighborhood of each oxygen atom of a water molecule is extracted and transformed into a power spectrum using SOAP method. Following standard preprocessing, this vector is fed into a variational autoencoder (VAE) model, comprising fully connected layers that construct encoder and decoder blocks.}
\label{figure:fig1}
\end{figure}

The VAE architecture used in IceCoder features encoder and decoder blocks composed of fully connected layers, each activated by the Gaussian Error Linear Unit (GeLU) function \cite{GELU}. The encoder comprises four hidden layers with 1500, 600, 400, and 100 nodes, respectively, leading to a bottleneck layer with 2 nodes, and the decoder block mirrors the encoder. The hyperparameters, including the number of nodes and layers, were meticulously tuned using Optuna \cite{optuna_2019} to balance model performance and computational efficiency.

For the SOAP representation, we set the radius of the atomic environment to \(r_c = 10\) Å, ensuring comprehensive capture of the neighboring atomic network. The choice of \(r_c\) is crucial, as it significantly impacts the model's ability to preserve relevant patterns in the spectra. The SOAP descriptor was expanded up to \(l_{\text{max}} = 6\) and \(n_{\text{max}} = 8\), with parameters like \(\sigma = 0.25\) Å carefully selected to optimize the capture of neighborhood information. The DScribe package was employed to construct these descriptors \cite{Himanen2020}.

During training, the weights and biases of each neuron in the hidden layers of VAE are adjusted, starting with random initialization. The processed SOAP vector, a 952-dimensional vector in this case, is fed into the model, as illustrated in Figure~\ref{figure:fig1}. The model is trained using the Adam optimizer \cite{adam} with a learning rate of 0.00001. Mini-batch learning is employed with a batch size of 128, adjusting the weights and biases with each iteration through the batch. To avoid overfitting, the early stopping method has been implemented.

To generate ideal lattice structures at \(T = 0K\), we utilized the GenIce package \cite{Matsumoto2017}. The energy minimization of the corresponding phases was achieved through a combination of steepest descent and the conjugate gradient method. A comprehensive dataset of water molecules was constructed using molecular dynamics simulations with the TIP4P/Ice water model \cite{Bore2022}. The TIP4P/Ice model, a refined version of the 4-site TIP4P model, is specifically designed to construct phase diagrams of ice and amorphous water near the freezing point. It has been used in various applications, including predictions of homogeneous nucleation rates, the binding free energy of antifreeze proteins on the ice interface, and the shear viscosity of ice \cite{Espinosa2014, HajiAkbari2015, Mochizuki2018, Louden2018}.

In simulation, Van der Waals interactions were calculated using a cutoff method with a 1.0 nm radius, while Coulomb interactions were handled using the Particle Mesh Ewald (PME) algorithm with a 1.0 nm short-range cutoff \cite{Essmann1995}. Newton's equations of motion were solved using the leap-frog algorithm with a 2 fs time step. All covalent bonds were constrained using the LINCS algorithm \cite{Hess1997}, and Lennard-Jones parameters for various atom types were calculated using the Lorentz-Berthelot combination rule \cite{DELHOMMELLE2001}. The GROMACS package, version 2019.6, was used for all simulations \cite{Abascal2005, Abascal2007, Jorgensen1988, Jorgensen1996, gmx1, Abraham2015, Berendsen1995}.

Following energy minimization via the steepest descent algorithm, equilibration and production simulations were performed under the NPT ensemble. System-specific temperature and pressure values from the TIP4P/Ice model’s phase diagram were selected to ensure stability of those specific ice phases in MD simulations. Only homogeneous phases make up the dataset that was produced to train the model. Table~\ref{tab:table1} presents the particular thermodynamic conditions that were used to simulate the phases.

For these systems, a 1-ns equilibration phase was followed by a 5-ns production simulation  during which potential energy was monitored to detect any phase transitions. Velocity-rescale thermostat \cite{Bussi2007} and  Parrinello-Rahman barostat\cite{Parrinello1981} were used to maintain the temperature and pressure of the systems, respectively.

\begin{table}
\caption{\label{tab:table1} Thermodynamic conditions used in molecular simulation of ice-liquid phases}
\begin{tabular}{ccc}
\hline
Phase & Temperature (K) & Pressure (bar)\\
\hline
Ice-Ih&230&1\\
Ice-Ic&230&1\\
Ice-II&230&3000\\
Ice-III&230&3000\\
Ice-V&230&6000\\
Ice-VI&230&8000\\
Ice-VII&230&50000\\
Ice-XII&250&1000\\
Ice-XVII&230&0.5\\
\hline
\end{tabular}
\end{table}

For the test cases, we have prepared two systems featuring coexisting phases: (i) ice-Ih/liquid and (ii) ice-Ih/ice-Ic/liquid configurations. To simulate hexagonal ice growth in the ice-Ih/liquid system, we constructed an ice/liquid interface by placing an ice-Ih crystal slab consisting of 2000 water molecules adjacent to 2441 liquid water molecules within a simulation box measuring 3.93 x 3.69 x 10.04 nm³. The system was maintained at a temperature of 265K using a velocity-rescale thermostat \cite{Bussi2007} and at a pressure of 1 bar using a Parrinello-Rahman barostat \cite{Parrinello1981}. A 100 ns production run was conducted to observe the complete growth of ice.

For the system featuring the coexistence of cubic and hexagonal ice at 265K and 1 bar, crystal slabs of similar size were placed within a larger simulation box measuring 7.05 x 5.13 x 11.72 nm³, surrounded by bulk water. During energy minimization and equilibration, the positions of the ice slabs were restrained. Over time, the system evolved, and after a 500 ns production run, it resulted in a fully cubic ice configuration.

\section{Results and Discussion}
Each ice phase exhibits unique geometric configurations, resulting from different arrangements or orientations of neighboring water molecules around a central reference molecule. This diversity in ice-phase structures highlights the complexity of their spatial organization that creates complexity in simulating ice/liquid phases. Some representative snapshots of the unit cells of the ice phases are shown in Figure~\ref{fig:intro}(a). Their mutual structural deviation is clear from the plot of the radial distribution function of oxygen-oxygen atoms of the water molecule in their corresponding phase(Figure~\ref{fig:intro} (b)).

\subsection{Ice phases in classical OP space}
Classical order parameters are typically designed using intuition about the specific problem at hand. Often, they focus on the arrangement of neighboring atoms within the local environment, encoding this geometric information through appropriate mathematical functions of the atomic coordinates. We have first tested several such standard OPs in terms of their ability to distinguish between the ice phases under consideration.

\begin{figure}
	\centering
	\includegraphics*[width=.95\linewidth]{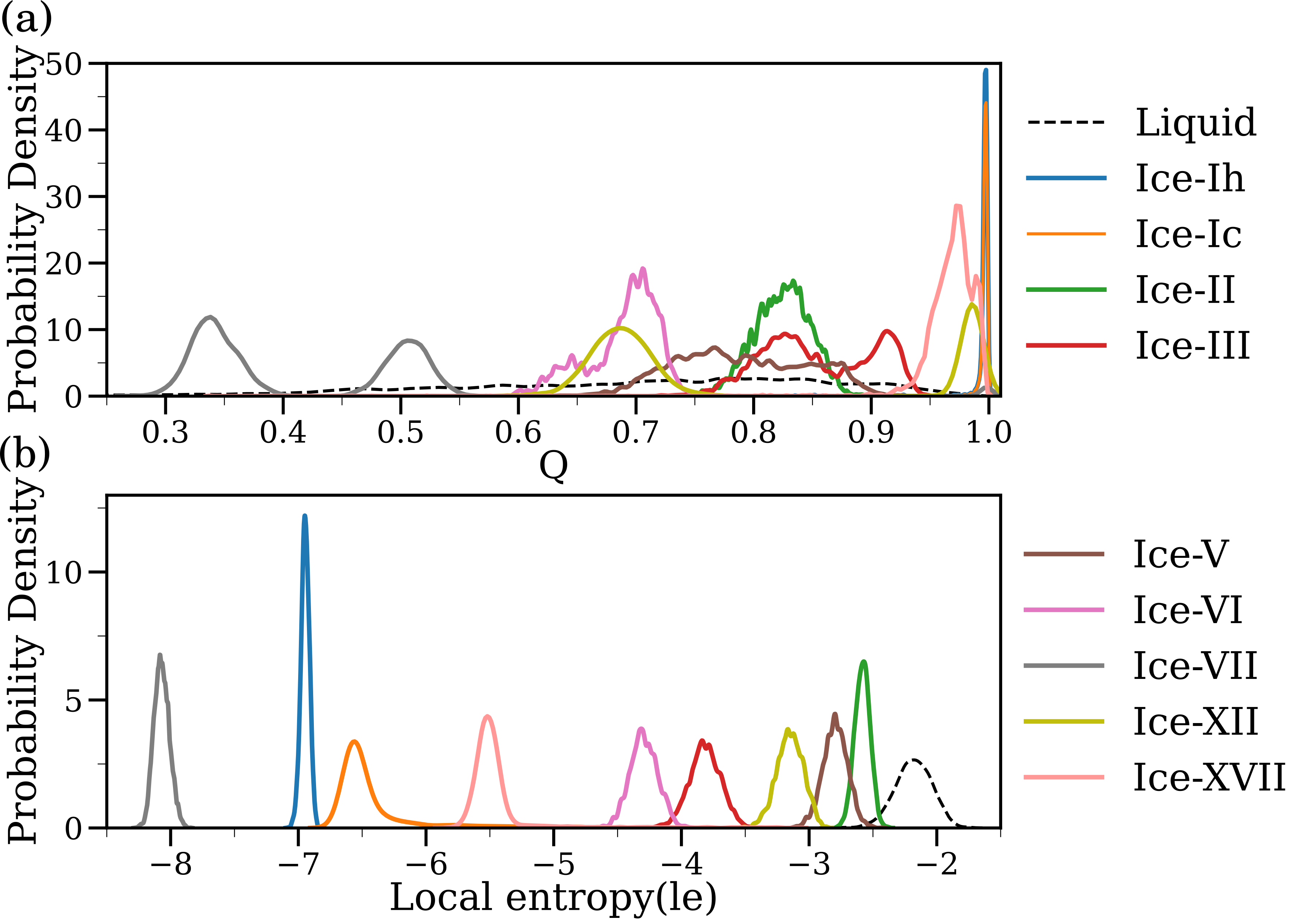}
	\caption{Probability distribution of order parameters for ice phases, (a) orientational tetrahedral order parameter(Q) and (b) local entropy fingerprints(le).}
	\label{figure:fig2}
\end{figure}

Tetrahedral order parameter considers tetrahedral ordering in the hydrogen bonding network of neighboring water molecules of the oxygen molecule of interest. It yields a value of one for a perfect tetrahedral orientation and zero for a completely disordered arrangement, as seen in the liquid state of water. We extracted frames from the simulated trajectory and energy-minimized them. When plotting the probability density of the tetrahedral order parameter for each energy-minimized ice phase structure, we observed significant overlap(Figure~\ref{figure:fig2} (a)). This overlap is expected to increase considerably when thermal fluctuations are taken into account.

We conducted a similar analysis for all other order parameters as well. In the literature, people have used the third and fourth variant of Steinhardt's order parameter to dissect this problem, from its success in resolving regular crystal space (hcp, bcc, hcp and sc). The joint space of $q_3-q_6$ is able to separate ice-Ih, ice-Ic, Ice-VII from all other states, but there remains overlap in the high-pressure ice phases, and worst they fall in the liquid region(Figure~\ref{fig:intro} (c).  Entropy fingerprint, specifically its local variant, namely the local entropy, seems to be separating a significant number of ice phases (energy minimized structure)(Figure ~\ref{figure:fig2} (b)). Although the notion of boundary diminishes in the presence of thermal fluctuation, aka for simulated trajectory data.

The boundaries of individual phases in these order parameter spaces are sensitive to thermal fluctuations and often they overlap under such scenarios. Even worse, the liquid phase usually spans over other ice phases. From the analysis of traditional order parameters, we can safely say that typical order parameters fail to adequately categorize many ice phases. Although hexagonal ice (Ice-Ih), cubic ice (Ice-Ic), Ice-VII, and Ice-XII are easily distinguishable within those order parameter space, numerous other ice phases overlap, posing a challenge for classification.

\subsection{Ice phases in VAE latent space} 
We have analyzed the atomic environments of oxygen atoms in the ice phases we simulated, transforming these environments into roto-translational and permutation-invariant SOAP vectors. In our case, each SOAP vector, representing the atomic neighborhood of an oxygen atom, is 952-dimensional. Following extraction, we apply MinMaxScaler to normalize the elements of vectors to a range between 0 and 1. This preprocessed data was subsequently used as input features to our Variational Autoencoder (VAE) model.

\begin{figure}
	\centering
	\includegraphics*[width=.95\linewidth]{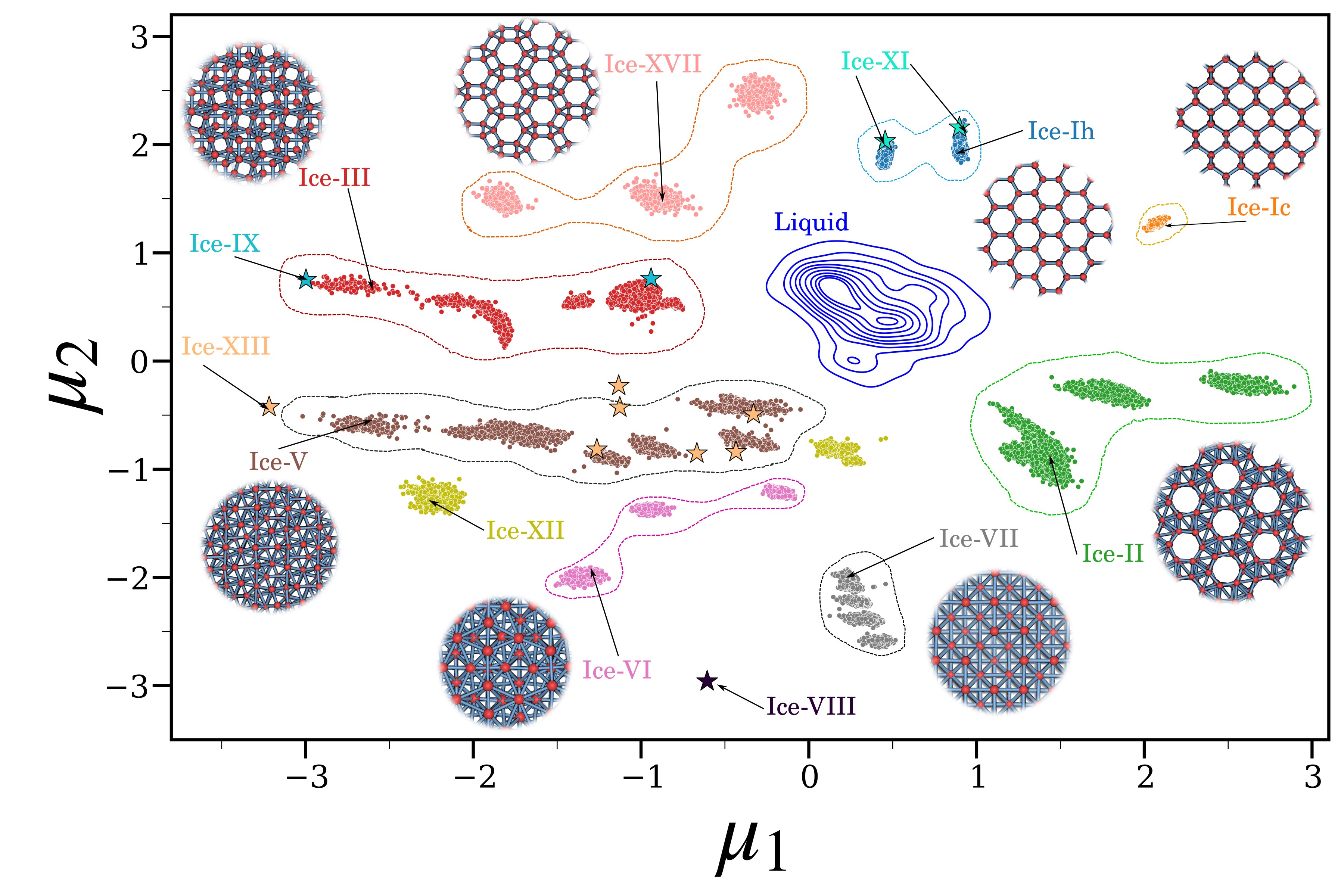}
	\caption{Distinctive positions of ice phases in the VAE latent space. The approximate boundaries between phases are indicated by dotted lines. Star symbols represent the hydrogen-ordered counterparts of the ice phases(in a pairwise manner, Ice-XI-Ice-Ih, Ice-IX-Ice-III, Ice-XIII-Ice-V, and Ice-VIII-Ice-VII). A representative snapshot of few phases is provided within the plot for visual reference.}
	\label{figure:fig3}
\end{figure}

Once the VAE model was trained, we utilized it to project the atomic environments of the respective ice phases onto the lower-dimensional latent space (2D in this instance). When we projected the SOAP vectors from energy-minimized ice structures into the latent space, we observed distinct separations among the ice phases, with the exception of Ice-XII (Figure~\ref{figure:fig3}). Notably, hydrogen-ordered and hydrogen-disordered pairs occupied the same regions, attributed to the static orientation of oxygen molecules, with primary differences arising from the positioning of hydrogen atoms, which are largely similar in the higher-dimensional input space. However, the distribution for the hydrogen-ordered phase was considerably narrower compared to the hydrogen-disordered phase, as illustrated in Figure~\ref{figure:fig3}, where the hydrogen-ordered counterparts of the corresponding ice phases are marked with star symbols.

\begin{figure*}
	\centering
	\includegraphics*[width=.95\linewidth]{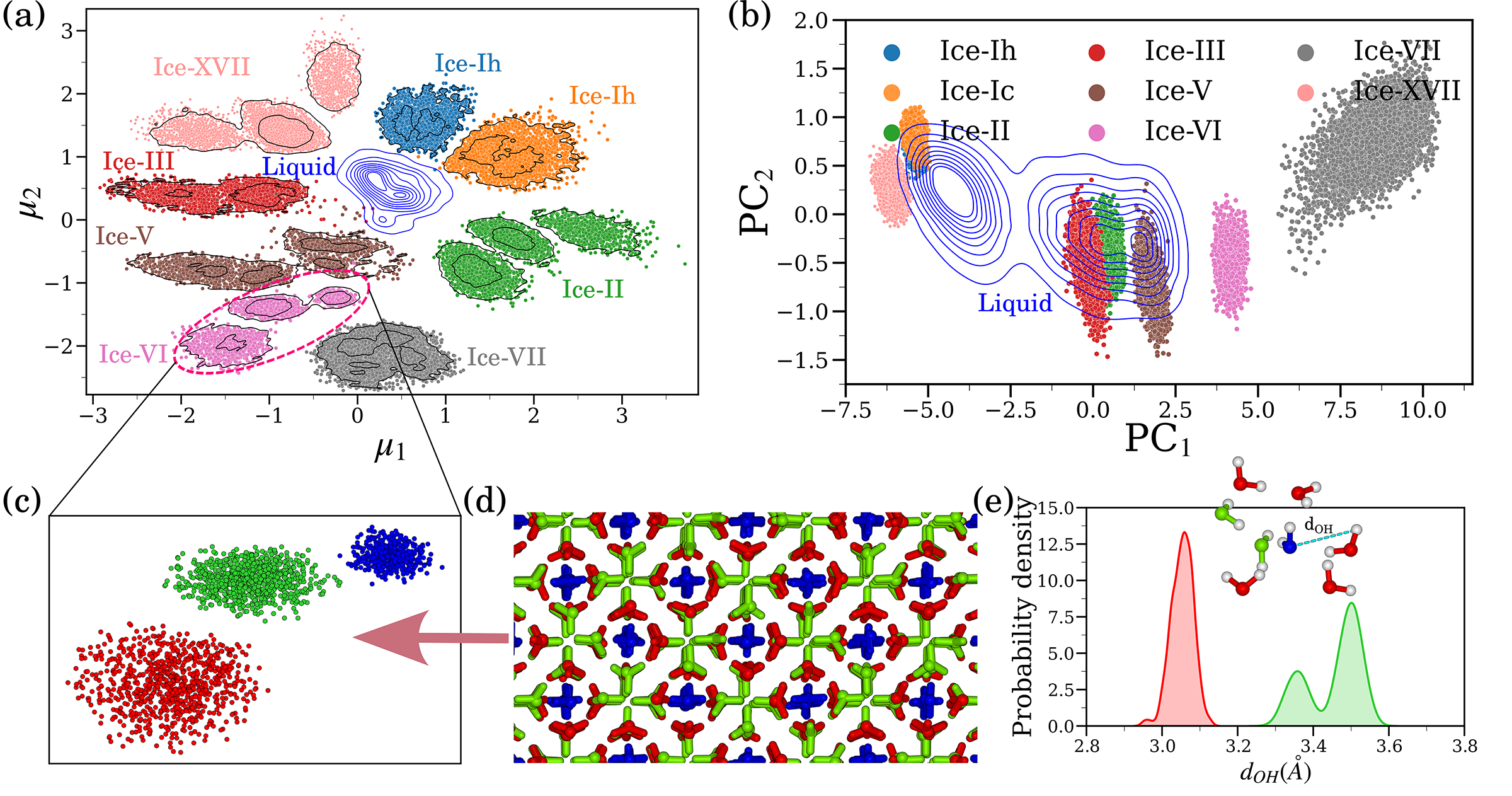}
	\caption{(a) Oxygen atoms from both ice and liquid phases in the MD simulation are projected into the mu1-mu2 space, with their colors corresponding to the text for visual clarity. (b)The same SOAP data is projected onto the top two principal components($PC_1-PC_2$) of PCA, revealing significant overlap between the ice phases, with the liquid water region entirely engulfing some of them. (c)-(e) demonstrate the resolution of this approach within individual phases. (c) Three distinct clusters within Ice-VI are highlighted in red, green, and blue for visual differentiation. (d) A ball-and-stick model visually represents the VAE-predicted clusters within the Ice-VI crystal lattice, with each atom colored according to its cluster assignment. (e) This figure emphasizes the variation in distances between the central oxygen atom (colored blue) and the farthest hydrogen atoms (colored lime green and red) of water molecules, illustrating the differences between molecules in distinct environments as indicated by the color scheme.}
	\label{figure:fig4}
\end{figure*}

\begin{figure}[H]
	\centering
	\includegraphics*[width=.85\linewidth]{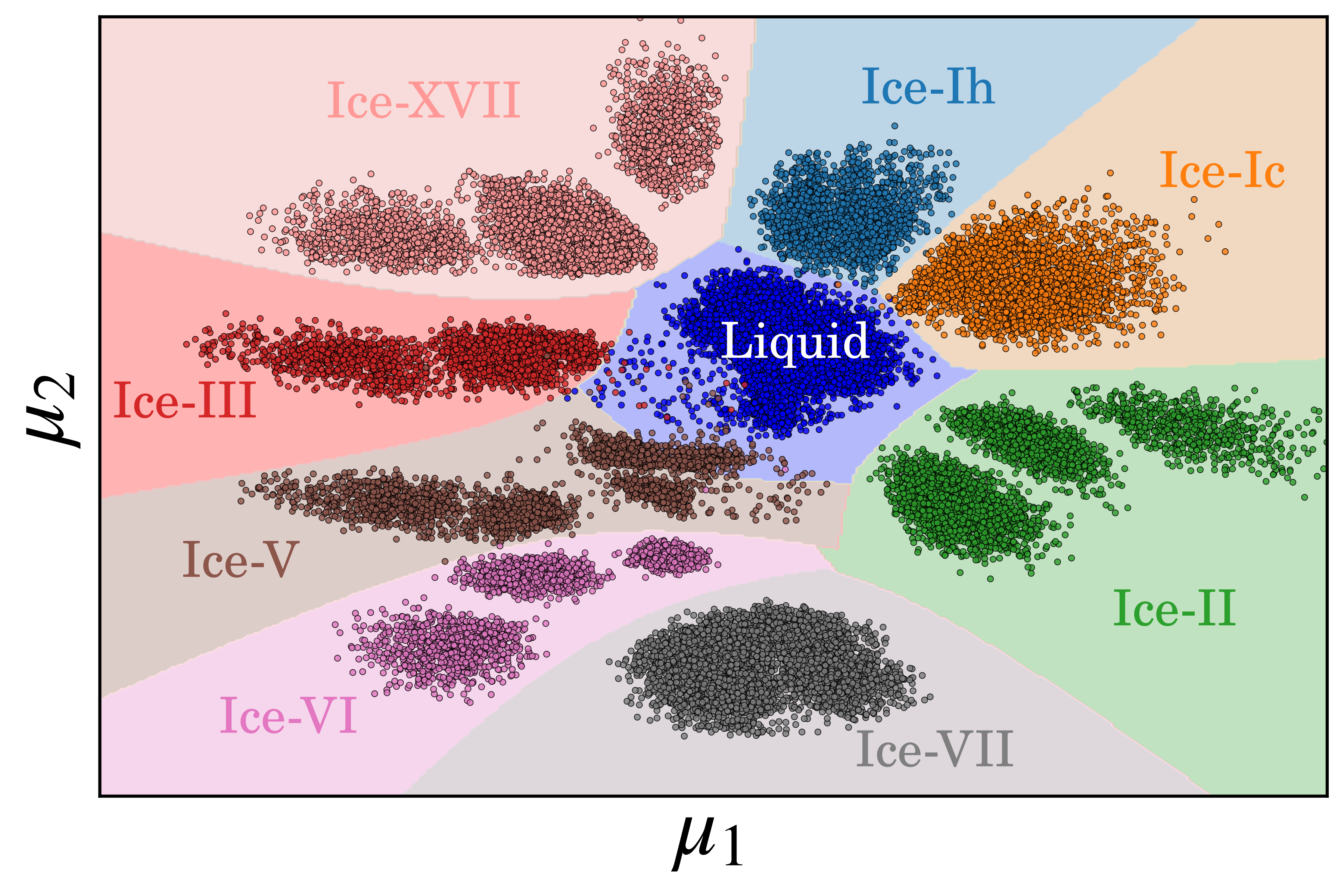}
	\caption{Support vector machine used to specify the boundaries of the phases. It was used after to predict the unknown phases present at simulation time. The name of the phased are written in the regions for clarification.}
	\label{fig:svm}
\end{figure}

A key question is whether the separation of these phase boundaries persists during molecular dynamics (MD) simulations. Is this approach sensitive enough to detect thermal fluctuations or the phases overlap?  To find an answer, we projected data from MD simulation trajectories using the same trained model. Figure~\ref{figure:fig4}(a) shows eight ice phases simulated with the TIP4P/Ice force field at thermodynamic equilibrium (details of the simulation and system are provided in the Methods section). Distinct boundaries between the phases are evident, showcasing the strength and robustness of this approach.
For practical applications, defining these phase boundaries in an automated manner is crucial for predicting or classifying ice phases. Support Vector Machines (SVM)\cite{Cortes1995}  are commonly used for this purpose due to their ability to handle non-linear boundaries with robustness. In our study, we employed an SVM with a radial basis function (RBF) kernel to separate the $\mu_1-\mu_2$ space. The trained SVM can predict the ice phase in terms of labels and is also used in later sections to detect local ice phases. Figure~\ref{fig:svm} illustrates the boundaries defined by the SVM, based on the MD simulation data projected into the $\mu_1-\mu_2$ space of the VAE.

\subsection{Intra-phase resolution in latent space}
The combination of the variational autoencoder (VAE) model with SOAP not only effectively classifies the ice phases but also discerns the subtle differences within each phase. As illustrated in Figure~\ref{figure:fig4}(a) and (c), most ice phases are divided into subclusters, with each subcluster representing a unique atomic environment corresponding to the respective ice phase. For instance, in the case of ice-VI, three distinct clusters within the same phase are observed in the $\mu_1-\mu_2$ space. When examining ice-VI through the lens of VAE prediction, we notice distinctive atomic neighborhoods surrounding the central oxygen atom (depicted in blue), with variations in orientation even among visually similar molecules colored in red and green(Figure~\ref{figure:fig4}(d)). 
To verify that these VAE-predicted clusters differ geometrically, we performed a distance analysis. For each VAE-identified cluster, we measured the distance from the central oxygen atom to the farthest hydrogen atoms. A schematic of this distance is shown in the inset of Figure~\ref{figure:fig4}(e). Figure~\ref{figure:fig4}(e) plots the probability density distribution of this parameter for two Ice-VI sub-clusters(red and green), revealing clear differences between them. This highlights the model's ability to detect fine-grained differences in atomic arrangements, showcasing the high resolution of the VAE approach.

\subsection{Non-linear vs linear dimensionality reduction method}

The smooth overlap of atomic positions (SOAP) effectively captures unique atomic environments. However, this raises the question: is a non-linear dimensionality reduction technique like VAE truly necessary to project these phases into a lower-dimensional space, or could a simpler linear transformation, such as principal component analysis (PCA), suffice? To evaluate the need for VAE, we compared its projection with that of PCA for the ice phases in the PC1-PC2 space. As expected, the PC1-PC2 projection shows significant overlap between the ice phases, with liquid water notably spreading over the ice phases, perhaps due to the non-linear manifold present in the input data, which is undesirable. This underscores the value of using VAE for distinguishing between ice phases. A similar result was found by Monserrat et al., where PCA projection of SOAP data also showed ice phases being engulfed by the water region\cite{Monserrat2020}. Nevertheless, some ice phases do occupy distinct positions in the PC1-PC2 space, suggesting that linear methods may be useful in specific cases. This comparison is illustrated in Figure~\ref{figure:fig4}(b).

\subsection{Case study I: Capturing heterogeneity and transition in a 2-phase system}
\begin{figure}
	\centering
	\includegraphics*[width=.95\linewidth]{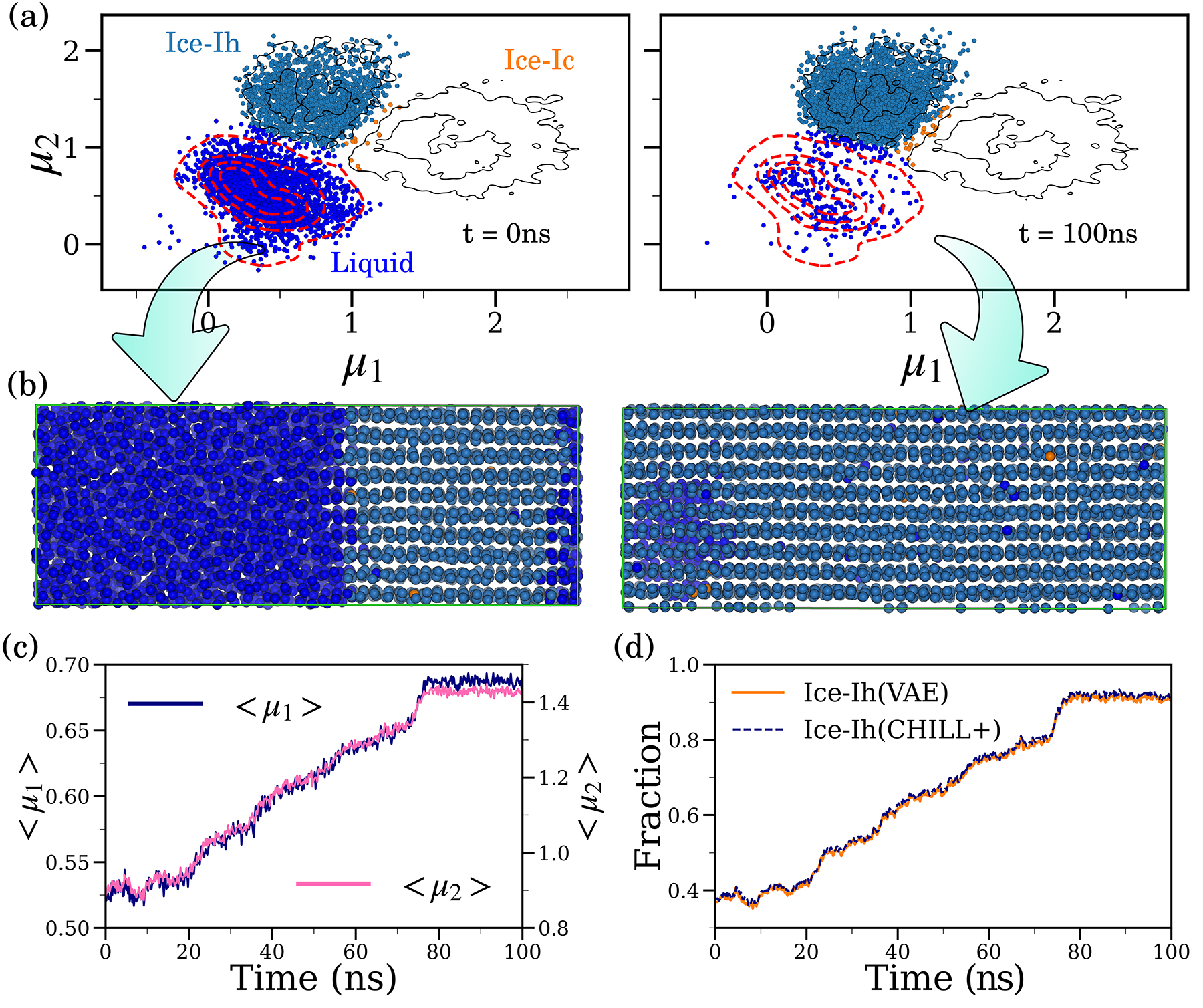}
	\caption{Growth of Hexagonal Ice in an Ice-Water Mixed System. (a) Oxygen atom of water molecules are projected onto VAE space at $t = 0ns$ and $t = 100ns$. They are colored according to VAE prediction. (b) Snapshots provide a visual representation, with sphere color indicating corresponding phases. (c)Twin plot of time evolution of frame averaged latent space order parameters. They serve as global descriptor of the transition. (d) Time evolution of fraction of hexagonal ice molecules depicted for both VAE and $CHILL+$ methods, showing good agreement between algorithms.}
	\label{figure:fig5}
\end{figure}

Below the supercooled regime of water, spontaneous growth of ice-Ih can occur from a post-critical nucleus or ice surface\cite{Moore2011}. To simulate this, we prepared a system consisting of a mixture of hexagonal ice and liquid water. In the initial setup, an ice-Ih slab was introduced alongside bulk water to induce ice formation. Over the course of a 100 ns simulation, we observed the gradual expansion of hexagonal ice from the interface. This process is characterized by the time evolution of the average of $\bar{q}_4$ and $\bar{q}_6$ order parameter, as shown in Figure S4. The same phenomenon is verified by tracking the fraction of ice-Ih molecules formed, determined using the CHILL+ algorithm, presented in Figure~\ref{figure:fig5}(d).

We applied our pre-trained VAE model to SOAP data extracted for oxygen atoms from each frame of the trajectory, projecting them into the VAE’s latent space. Figure ~\ref{figure:fig5}(a) clearly shows that at $t=0 ns$, the system contained equal proportions of ice-Ih and liquid molecules, while by $t=100 ns$, most liquid molecules had transitioned to the ice-Ih region, confirming both visual observations and traditional analyses. Figure ~\ref{figure:fig5}(b) provides a visual representation of the system states at the above mentioned times, colored according to VAE predictions (following the previously established color scheme for ice and liquid phases).

When averaged on all molecules in a frame, the latent space variables, $\mu_1$ and$\mu_2$, serve as global order parameters for the system, correlating with traditional order parameters. Figure ~\ref{figure:fig5}(c) shows the time evolution of $\langle \mu_1 \rangle$ and $\langle \mu_2 \rangle$, providing a clear interpretation of the phase transition. Additionally, we used the pre-trained SVM to predict the number of Ice-Ih molecules at each time point, comparing these results with those obtained using the CHILL+ algorithm. The results show excellent agreement, as illustrated in Figure ~\ref{figure:fig5}(d), which presents a comparison between the two approaches.
\subsection{Case study II: Capturing heterogeneity and transition in a 3-phase system}
In another case study, we examined the coexistence of cubic and hexagonal ice within a liquid water environment under favorable thermodynamic conditions (see Methods for details). Initially, the system contained equal amounts of hexagonal and cubic ice (arranged as slabs) along with a layer of bulk water in between them. As the simulation progressed, the liquid water gradually transformed into both types of ice, causing the water layer between the ice slabs to thin. Eventually, the interface of the hexagonal ice came into contact with that of the cubic ice. Due to the stacking disorder in cubic ice, the interfacial tension between the two phases drove the system toward a cubic ice-dominated state within approximately 370 ns, resulting from the conversion of hexagonal ice into cubic ice.

\begin{figure}
	\centering
	\includegraphics*[width=.95\linewidth]{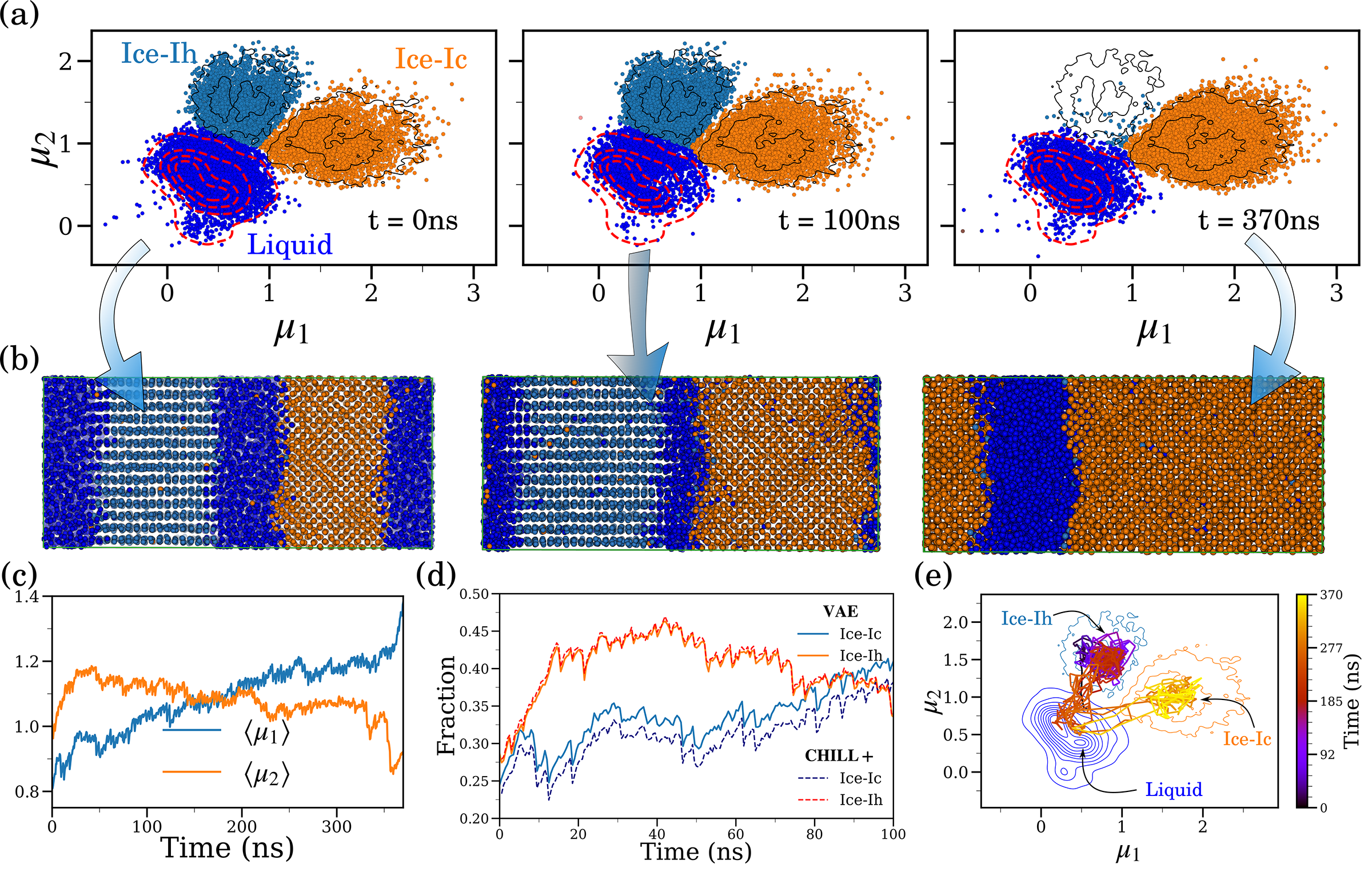}
	\caption{Case Study: Coexisting Phases of Cubic Ice and Hexagonal Ice with water at 250K and 1 bar. (a) VAE projection of the system at $t = 0ns$, $t = 100ns$, and $t = 370ns$.(b) Snapshots of the system at abovementioned time stamp, colored accordingly VAE predictions. (c) Time evolution of frame averaged latent space order parameters for the simulation. The averaged order parameter serves as a global descriptor of the phase transition. (d) Time(ns) vs fraction of number of ice molecules. Solid line from VAE prediction and dashed line for the $CHILL+$ algorithm, previous color convention for the ice phase is followed. (e) A molecular-level transition is visualized in the VAE space. Color of the line corresponds to the time step of the simulation which is mapped to the colorbar given.}
	\label{figure:fig6}
\end{figure}

These events were consistently reflected in the $\bar{q}_4-\bar{q}_6$ space, as we plotted the time evolution of the averaged values of these order parameters(Figure S5), as well as CHILL+ predictions(Figure ~\ref{figure:fig6}(d)), similar to Case I. Figure ~\ref{figure:fig6}(a) shows the VAE projection of the system at $t=0 ns$, $t=100 ns$ and $t=370 ns$. The plots show that the water layer between the ice phases thins gradually, and by around 100 ns, the interfaces of the two phases come into contact. Subsequently, the system transitions into a cubic ice-dominated state, where the hexagonal ice melts and converts into cubic ice.

Figure ~\ref{figure:fig6}(b) provides a visual representation of the system at the aforementioned time points, with colors indicating VAE predictions. In Figure ~\ref{figure:fig6}(c), we present the plot of time vs frame-averaged $\mu_1-\mu_2$ order parameters, which clearly demonstrate the crossover of the phase transition. The averaged $\mu_1$ correlates with the number of cubic ice molecules, showing a steady increase over time, while $\langle\mu_2\rangle$ correlates with number of hexagonal ice, showing a crossover. A correlation map between the number of ice molecules and the frame-averaged $\mu_1-\mu_2$ values is shown in Figure S7.

Figure ~\ref{figure:fig6}(d) compares the number of different ice molecules predicted by the SVM, based on the VAE latent representation of the oxygen atoms, with those predicted by CHILL+. Both approaches show excellent agreement. Finally, Figure ~\ref{figure:fig6}(e) demonstrates that our method can probe transitions at the single-molecule level. The figure shows an individual water molecule transitioning from hexagonal ice to liquid and then converting into cubic ice finally.

To assess the predictive accuracy of our framework, we compared its performance with CHILL+\cite{Nguyen2014}, a tool that detects hexagonal, cubic, and clathrates using geometric criteria. The strong agreement between the results obtained from CHILL+ and those predicted by our VAE framework underscores the accuracy and robustness of our approach, highlighting its effectiveness as a general and data-driven methodology with algorithm-specific accuracy.
If we go from a local to the global descriptor, frame-averaged latent space variable serves the same purpose. The events associated with transitions are well captured using those variables.

\section{Conclusion}
Designing a metric capable of distinguishing between all ice phases would be highly beneficial in molecular dynamics simulations, as it could serve as a direct probe to monitor phase transitions and other notable changes occurring within the system during simulation. Traditional methods tend to be overly case-specific and often approach the problem in a coarse-grained manner. The mathematical formulation of these order parameters has typically been tailored to specific states corresponding to particular transitions, limiting their general applicability. Similarly, data-centric methods often lose accuracy when confronted with the vast complexity of ice phases. Numerous approaches have been proposed to address this challenge\cite{Fulford2019,Kim2020}.

Our approach merges physical intuition with the strengths of autoencoders to extract meaningful representations of the local neighborhood of ice molecules, which are distinctive across most ice phases. This method is capable of differentiating between various ice states and identifying different heterogeneities within the same crystal. However, certain challenges remain. Specifically, for order-disorder transitions in ice pairs (conjugates) where the primary difference lies in hydrogen ordering, this approach struggles. In such cases, the only reliable method of distinguishing these states is through analysis of their dynamics or flexibility.

\section{Data and Software Availability}
The data that support the ﬁndings of this study are available from the corresponding author on a reasonable request. Python codes and tutorial for running the IceCoder software are available for public use at https://github.com/teamsuman/IceCoder.

\begin{acknowledgement}
The authors thank the Technical Research Centre (TRC) computing facilities of S. N. Bose National Centre for Basic Sciences (SNBNCBS), established under the TRC project of the Department of Science and Technology (DST), Govt. of India. D.M. thanks SNBNCBS for the fellowship.
\end{acknowledgement}

\bibliography{reference}

\end{document}